\documentclass[%
superscriptaddress,
preprint,
showpacs,
 amsmath,amssymb,
 aps,
pre
]{revtex4-1}

\usepackage[shortlabels]{enumitem}
\usepackage{graphicx}
\usepackage{dcolumn}
\usepackage{bm}
\usepackage{hyperref}

\allowdisplaybreaks 

\begin{document}


\title{Intelligent buses in a loop service: Emergence of \emph{no-boarding} and \emph{holding} strategies}

\author{Vee-Liem Saw}
\email{Vee-Liem@ntu.edu.sg}
\affiliation{Division of Physics and Applied Physics, School of Physical and Mathematical Sciences, 21 Nanyang Link, Nanyang Technological University, Singapore 637371}
\affiliation{Data Science and Artificial Intelligence Research Centre, Block N4 \#02a-32, Nanyang Avenue, Nanyang Technological University, Singapore 639798}
\author{Luca Vismara}
\email{vism0001@e.ntu.edu.sg}
\affiliation{Complexity Institute, Interdisciplinary Graduate Programme, 61 Nanyang Drive, Nanyang Technological University, Singapore 637335}
\affiliation{Division of Physics and Applied Physics, School of Physical and Mathematical Sciences, 21 Nanyang Link, Nanyang Technological University, Singapore 637371}
\author{Lock Yue Chew}
\email{lockyue@ntu.edu.sg}
\affiliation{Division of Physics and Applied Physics, School of Physical and Mathematical Sciences, 21 Nanyang Link, Nanyang Technological University, Singapore 637371}
\affiliation{Data Science and Artificial Intelligence Research Centre, Block N4 \#02a-32, Nanyang Avenue, Nanyang Technological University, Singapore 639798}
\affiliation{Complexity Institute, 61 Nanyang Drive, Nanyang Technological University, Singapore 637335}
%

\date{\today}

\begin{abstract}
We study how $N$ intelligent buses serving a loop of $M$ bus stops learn a \emph{no-boarding strategy} and a \emph{holding strategy} by reinforcement learning. The high level no-boarding and holding strategies emerge from the low level actions of \emph{stay} or \emph{leave} when a bus is at a bus stop and everyone who wishes to alight has done so. A reward that encourages the buses to strive towards a staggered phase difference amongst them whilst picking up people allows the reinforcement learning process to converge to an optimal Q-table within a reasonable amount of simulation time. It is remarkable that this emergent behaviour of intelligent buses turns out to minimise the average waiting time of commuters, in various setups where buses have identical natural frequency, or different natural frequencies during busy as well as lull periods. Cooperative actions are also observed, e.g. the buses learn to \emph{unbunch}.
\end{abstract}

\maketitle


\section{Introduction}

Consider $N$ buses serving $M$ bus stops in a loop. These $M$ bus stops are staggered around the loop, each having a people arrival rate of $s$ people per second. Each of the $N$ buses has natural (angular) frequencies $\omega_1>\omega_2>\cdots>\omega_N$, respectively (excluding any time stopped at bus stops). When a bus arrives at a bus stop, it first allows people who wish to alight to do so, and then allows people to board. The alighting and boarding rate is $l$ people per second. Overall, the quantity $k:=s/l$ is a parameter that describes the level of demand for service. We assume that people from a bus stop would like to travel to the bus stop antipodally opposite to it, or the one just before travelling half a loop if $M$ is odd. By an analytical calculation, Ref. \cite{Vee2019} has shown that there is a critical $k_c(N)$:
\begin{align}\label{kc}
k_c(N)=\frac{1}{M}\sum_{i=1}^{N-1}\left(1-\frac{\omega_N}{\omega_i}\right),
\end{align} 
where all $N$ buses would be completely bunched into a single unit (i.e. completely synchronised) if $k>k_c(N)$. Also, we have the relations $\omega_i=2\pi f_i=2\pi/T_i$ between angular frequency, frequency, and period. [Note: Ref.\ \cite{Vee2019} assumed that boarding and alighting occur \emph{simultaneously} via different doors. Here, we assume that these occur \emph{sequentially} through one door, alighting followed by boarding. Thus, we should include an overall factor of $1/2$ into Eq.\ (\ref{kc}) for this paper, since the processes being sequential via one door instead of two different doors would double the time a bus dwells at a bus stop.] On top of that, numerical simulations with parameters based on values measured from a real university shuttle bus loop service showed that the buses are not persistently bunched (no phase-locking, completely unsynchronised) if $k<\bar{k}$, where $\bar{k}:=k_c(2)$ is the corresponding system with $N=2$ buses having natural frequencies $\omega_1$ and $\omega_N$. The bus system would have some buses persistently bunched (partial synchronisation) if $\bar{k}<k<k_c(N)$. In the case where all buses have identical natural frequency, all buses would typically end up bunching into a single unit unless $k$ is sufficiently low such that each bus only spends the minimum amount of time stopping at each bus stop. Therefore, bus bunching is a perennial phenomenon, and it is of great interest to employ strategies such that the buses are able to maintain a regular headway between them, always remaining staggered.

A common strategy that has been widely studied is the \emph{holding strategy} \cite{Abk84,Ros98,Eber01,Hick01,Fu02,Bin06,Daganzo09,Cor10,Cats11,Gers11,Bart12,Chen15,Moreira16,Wang18}: If a bus is too fast, it would exercise an extended stoppage duration to correct for the headway from the bus in front of it --- otherwise it would bunch with it. Holding back buses however, may tend to slow down the system and would require that some slack in the schedule has been allocated beforehand. Recent studies explore the opposite, viz. a \emph{no-boarding strategy} \cite{Vee2019b,Vee2019c}: a slow bus would always allow passengers to alight at a bus stop, but would disallow boarding and leave the bus stop if it is \emph{too slow} in order to speed it up. Generally, the no-boarding strategy works well for a bus system with identical natural frequency by maintaining the buses' headways close to being ideally staggered. If the bus system has frequency detuning (i.e. the natural frequencies are $\omega_1>\omega_2>\cdots>\omega_N$) the no-boarding strategy is also successful during the busy period when $k$ is high since there is enough demand to slow down the ``faster bus''. Surprisingly, this strategy backfires during the lull period when $k$ is low, as the slow bus has been sped up to the maximum by picking up nobody whilst there is insufficient demand to slow down the fast bus enough. Consequently, the system is effectively operating with one less bus (since the slowest bus is almost always disallowing boarding).

The purpose of this paper is to explore if there are ways beyond what was analytically studied in Ref.\ \cite{Vee2019b}, such that a no-boarding policy may actually be salutary, especially in the lull period for a bus system with frequency detuning. Investigating a bus system with frequency detuning is crucial, because human-driven buses tend to move with different natural frequencies due to differing driving styles \cite{Vee2019}. Instead of implementing a human-thought-out or human-defined idea, we let buses figure out an ideal strategy via \emph{reinforcement learning} \cite{Sutton}: buses are given two actions whenever they are at a bus stop, viz. \emph{stay} or \emph{leave}, with no prejudice nor human input, other than a feedback on whether the average waiting time of the commuters is minimised. A so-called normal bus would \emph{stay} if there is somebody who wants to board, and \emph{leave} if there is nobody there. A no-boarding strategy would correspond to the bus deciding to \emph{leave} even if there is somebody who wants to board. Apart from the no-boarding strategy, the framework that is developed here would also be well suited to study the holding strategy: if there is nobody at the bus stop, a normal bus would \emph{leave}, but a holding strategy would correspond to \emph{stay}. By reinforcement learning (we shall employ Q-learning in this paper), the buses are initialised with random choices to execute. They would then progressively explore and converge to an optimal strategy, as is implied by the theory of Q-learning based on a Markov decision process \cite{Sutton}.

Applications of reinforcement learning to bus systems have been carried out for some forms of the holding strategy as well as with real-time data \cite{Eber01,Fu02,Bin06,Cor10,Moreira16}. However, as the no-boarding strategy appears to be a recent analytical investigation \cite{Vee2019b,Vee2019c}, there does not seem to be exclusive applications of reinforcement learning to bus systems with the no-boarding strategy, as well as a combination of holding and no-boarding strategies. Our consideration here with a single loop of $M=12$ bus stops is modelled after a university campus shuttle bus service that serves tens of thousands of students, staff and faculty members \cite{NTUnews,NTUautobuses}. This is thus a realistic system which also exists in many bus systems worldwide with loop services.

\subsection{Reward for reinforcement learning of the bus loop system}

With the goal of minimising the average waiting time of commuters for a bus to arrive at a bus stop, each time a bus is at a bus stop (and people who want to alight have done so), it executes either \emph{stay} or \emph{leave} and then receives the waiting time of the person ahead of the queue to board the bus (or who is supposed to board, but denied boarding if the bus leaves). However, we find that this feedback is problematic: the waiting time of each person has high variance. For instance, the luckiest person who arrives at the bus stop when a bus is there has zero waiting time, whilst the unluckiest person who arrives when a bus has just left would have maximum waiting time, with the other people's waiting times distributed between these two extremes. We have tried this direct feedback and found that the system almost never converges to a useful strategy due to the high variance in the feedback, with the average waiting time typically skyrocketing.

Alternatively, we note the following property for a bus loop system \cite{Vee2019b}: If the buses are staggered, then the average waiting time of commuters for a bus to arrive at the bus stop is minimised. The loop can be isometrically (i.e. preserving distance) mapped to a unit circle, which has well-defined phase angles from $0^\circ$ to $360^\circ$. Thus, we consider the feedback for the buses' actions as \emph{the phase difference between itself and the bus immediately behind it, $\Delta\theta$}. This phase difference experiences a more gradual change, thereby eliminating the high variance of measuring individual commuters' waiting times. The goal would be to keep this phase difference close to the staggered value. In other words, buses would be \emph{rewarded for being close to the staggered configuration}. For example, if there are two buses, then $\Delta\theta=0^\circ$ gives $0$ reward, whilst $\Delta\theta=180^\circ$ gives 1 point, with values in between scaling linearly. If $\Delta\theta>180^\circ$, then the reward linearly decreases to $0$ at $\Delta\theta=360^\circ$. Note however, this alone would lead to the buses striving to achieve the perfectly staggered configuration without any regard to the commuters, possibly even at the expense of not boarding anybody just to keep $\Delta\theta=180^\circ$. Therefore, we incentivise a reward of 1 point for each passenger who is picked up. A weighting hyperparameter can be selected such that the system is in a balanced region between closeness to staggered configuration and picking up people, where a bus aims to both pick up passengers and maintain a configuration that is nearly staggered. Of course, the choice and structure of the reward is arbitrary --- as long as it results in the intended minimisation of the average waiting time. This will be discussed in more detail in Section 3.

\subsection{Situations of interest}\label{sitofint}

The setup for the bus system undergoing Q-learning is as follows. Each bus has its own Q-table containing 72 states where they represent the phase difference as measured from the bus immediately behind it. This number of states is arbitrary, chosen to balance between not being too coarse and not taking too long for the simulations to run. Moreover for subsequent future applications on real-time non-stationary environments, it would be desirable for these buses to respond fast enough to adapt appropriately. Independent Q-tables allow different buses to possibly learn different strategies, where one bus may occasionally perform a ``sacrificial action'' for the system as a whole to benefit.

The 72 states would coarse grain the phase difference into bins of $5^\circ$. In each of these states, it records the two Q-values representing the expected total rewards for the two actions \emph{stay} or \emph{leave}, respectively. The buses typically move around on the road, where it must proceed with moving forward. When it reaches a bus stop, it must allow passengers who wish to alight to do so, i.e. we do not allow the possibility of stop-skipping. This is because we find it to be not beneficial to speed up the bus at the expense of another round of time spent on the bus for these passengers, or asking them to alight one stop earlier and ``walk their last mile'' to their intended destinations. Furthermore, this allows the reinforcement learning process to have better chances of converging to an optimal Q-table for every bus within a reasonable amount of simulation time.

The only time a bus is allowed to consider whether to execute \emph{stay} or \emph{leave} is when it is at a bus stop and there is nobody on the bus who wishes to alight. Here are the following situations that we would explore. A bus is allowed to consider its action when it is at a bus stop, and:
\begin{enumerate}
\item There is nobody to alight but there is somebody at the bus stop who wishes to board.
\item There is nobody to alight as well as nobody at the bus stop who wishes to board.
\item There is nobody to alight.
\end{enumerate}

The first situation is intended to create a possibility where the buses may learn to implement the no-boarding strategy, since it may learn to leave the bus stop even though there is somebody who wishes to board. The second situation is intended to create a possibility where the buses may learn to implement the holding strategy, since it may learn to stay at the bus stop even when there is nobody to pick up. Finally, the third situation allows the possibility for the buses to learn some combination of the no-boarding and holding strategies. In the first two situations, each bus has a Q-table with 72 states and each state contains two values --- one for \emph{stay} and one for \emph{leave}. For the third situation, there are 144 states because 72 states are when there is somebody who wants to board and another 72 states are when there is nobody who wants to board.

\subsection{Updating the Q-table}

In Q-learning \cite{Sutton}, when a bus is at a bus stop and has to pick an action $A$ of either \emph{stay} or \emph{leave}, it has to first determine what state $S$ it is presently in. To do so, it measures its phase difference $\Delta\theta$ with respect to the bus behind it. In this state, there are two actions and it chooses the one which has the highest Q-value --- unless it is in the $\varepsilon$-greedy exploration phase where there is a probability $\varepsilon$ of randomly selecting an action. According to the theory, after executing the action $A$ and receiving a reward of $R$, it should then subsequently measure again its phase difference from the bus immediately behind it to determine its future state $S'$ for the purpose of updating its Q-table with a future expected reward:
\begin{align}
Q(S,A)\leftarrow Q(S,A) + \alpha\left(R+\gamma\max_a\left(Q(S',a)\right)-Q(S,A)\right).
\end{align}
The hyperparameters $\alpha$ is the learning rate and $\gamma$ is the discount factor. The former determines how sensitively the Q-values would adjust due to new feedback, whilst the latter determines how seriously to believe an estimated future expected reward from its own Q-table.

However, since the bus only gets to execute an action at a bus stop when nobody wants to alight, if its action is \emph{leave}, then its future state would only occur quite some time later when it reaches a bus stop. We find that this has the effect of affecting reliable convergence as other things may happen during the time when this bus leaves the bus stop and reaches another bus stop: For example other buses would have picked up people at other bus stops and affect the overall waiting time of the commuters, as well as leading to a widely different future $\Delta\theta'$. To circumnavigate this issue in obtaining an estimated future expected reward for updating the Q-table, we impose that when a bus executes \emph{stay} in a state $S$ with phase difference $\Delta\theta$, then its future state $S'$ is defined as the same state with phase difference $\Delta\theta':=\Delta\theta$ since it remained; whilst if instead it executes \emph{leave}, then its future state $S'$ is defined as the state with phase difference $\Delta\theta':=\Delta\theta+5^\circ$, i.e. the phase difference has increased, since it moves to increase the phase difference from the bus behind it. Recall that this $5^\circ$ is the size of each state, since we use $72$ states to discretise the angles from $0^\circ$ to $360^\circ$.

\subsection{Bus system environment simulation parameters, reinforcement learning hyperparameters}

In all our simulations for the bus system environment, the parameters used are based on values measured from a real university shuttle bus loop service with $M=12$ bus stops \cite{Vee2019}. The value for the rate of people boarding/alighting is $l=1$ person per second. In the lull period, a representative average value for the people arrival rate at each bus stop is about $s=0.020$ people per second, whilst that in the busy period could be as high as $s=0.065$ people per second. The natural frequencies of the buses are measured to be in the range of 0.93 mHz to 1.39 mHz, or a natural period of $12$ minutes to $18$ minutes excluding time stopped at bus stops. We adapt these values accordingly in our simulations for the bus system environment. Each simulation time step corresponds to $1$ second.

For reinforcement learning, we carry out 1000 episodes, where each episode is 150 revolutions long. At the start of each new episode, the buses are randomly placed on the loop. The performance of the bus system in each episode is measured from the last 30 revolutions, where most of the transient part due to random initial conditions would have been weeded out. The system undergoes $\varepsilon$-greedy learning (i.e. there is a probability of $\varepsilon$ that a random action is taken), where $\varepsilon$ decays linearly from $1$ to $0.1$ in the first $200$ episodes, after which it remains at $0.1$ until the $700$th episode. The learning rate $\alpha$ is kept at $0.2$ for the first $700$ episodes. In the last $300$ episodes, we let the system fully exploit what they have learnt, with $\varepsilon=0$ and $\alpha$ toned down to $0.1$. The discount factor is always fixed at $\gamma=0.9$.

The first 200 episodes represent an \emph{exploration phase}, where the buses carry out many random actions due to the high value of $\varepsilon$. This is crucial to allow for the buses to avoid getting stuck in near-sighted local minima which may lead to missing out potentially better long-term strategies. The next 500 episodes form a mix of \emph{exploration and exploitation}, where the buses take advantage of their learned Q-tables but maintain some degree of exploration just in case they get stuck in some local minima. Finally, the last 300 episodes denote a \emph{fully exploitation phase}. Here, the buses still fine-tune their Q-tables since $\alpha=0.1$. The difference from previous episodes is that they now always take their best perceived action, never taking a random action anymore.

For each particular setup throughout this paper, we carry out (at least) five independent runs. Generally, we obtain essentially identical qualitative results when the same setup is repeated even though the learning process involves random initial conditions in each new episode and stochasticity in the $\varepsilon$-greedy exploration. This therefore assures robustness in our results.

Before diving into these interesting situations involving $N$ buses serving $M$ bus stops, we first consider the simplest or trivial situation of $N=1$ bus serving $M=12$ bus stops in the next section. With only one bus, there is no non-trivial phase difference with respect to another bus. Therefore, this bus must eventually \emph{learn to be a normal bus}, i.e. it \emph{stays} to pick up people when there is somebody who wishes to board and \emph{leaves} otherwise. In Section 3, we study the case of $N=2$ buses serving $M=12$ bus stops for each of the three situations described in Section \ref{sitofint}, followed by more buses in Section 4.

\section{\texorpdfstring{$N=1$}{N=1} bus learns to be a bus}

\begin{figure}
\centering
\includegraphics[width=16cm]{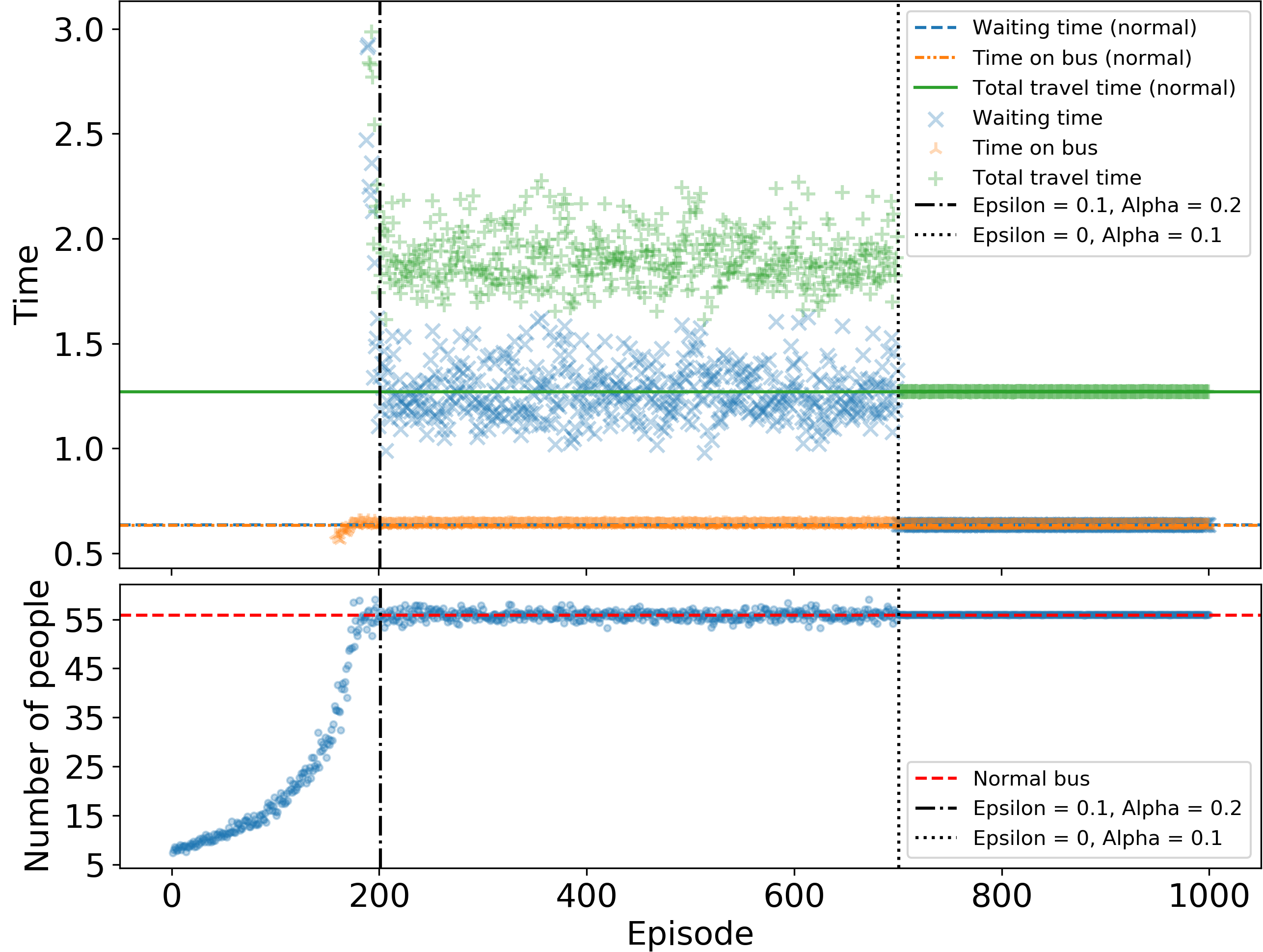}
\caption{A bus serving a loop of bus stops learns to be a bus by reinforcement learning.}
\label{fig1}
\end{figure}

With $N=1$ bus serving $M=12$ staggered bus stops in a loop, we aim to let this bus learn to be a bus, i.e. learn to \emph{stay} at the bus stop when somebody wants to board, and \emph{leave} when nobody is at the bus stop. Recall that by default, it must allow anybody who wishes to alight to do so. Fig.\ \ref{fig1} shows the average waiting time of commuters at the bus stop, average time spent on bus, average total travel time (which is the sum of waiting time and time spent on bus), and average number of people on the bus; all as functions of number of episodes. Since there is only one bus, its phase difference as measured from the bus behind it (itself) is always $360^\circ$ (or $0^\circ$). Hence, the reward is purely 1 point for each person picked up.

This reinforcement learning scenario corresponds to the third one listed in Section \ref{sitofint} where the bus decides on an action when it is at a bus stop and nobody wants to alight. The bus has a Q-table with two states, one when somebody wants to board and the other when nobody wants to board. Each state has two Q-values, one for \emph{stay} and one for \emph{leave}. This Q-table therefore has only four numbers. For the bus system environment, we set the natural period of the bus (excluding time stopped at bus stops) to be $T=12$ minutes, the rate of people arriving at each bus stop $s=0.010$ people per second. The unit of time for the top graph is $T=12$ minutes (this is the unit of time for all graphs in this paper, unless otherwise stated). 

As Fig.\ \ref{fig1} shows, the bus successfully learns to behave like a normal bus (\emph{stays} when there is somebody to board, \emph{leaves} when there is nobody to board), where it matches the performance of a hard-coded normal bus when it acts greedily in the last $300$ episodes based on the Q-values that it has learnt. In the $201$th to $700$th episode, since $\varepsilon=0.1$, it makes a random action once in every ten times, on average. A wrong action has ramifications on the waiting time of the commuters, since the bus leaves and they have to wait one additional revolution. Only when the bus acts greedily does the performance match that of a hard-coded normal bus. Nevertheless, the time spent on the bus is not too affected during the phase where $\varepsilon=0.1$, since passengers who want to alight must be allowed to do so. Large variance in the average number of people on the bus is observed before the $701$st episode due to the $\varepsilon$-greedy action selection. This variance vanishes in the last $300$ episodes when the bus acts greedily.

\section{\texorpdfstring{$N=2$}{N=2} buses learn no-boarding and holding}

Let us now study the interesting situations with $N=2$ buses serving a loop of $M=12$ staggered bus stops. We consider bus system environments with the following three setups throughout this section:
\begin{enumerate}[(a)]
\item Identical natural frequency, taking $T=12$ minutes to complete a loop (excluding time stopped at bus stops). The rate of people arriving at each bus stop is set at $s=0.010$ people per second.
\item Frequency detuning, with $T_1=12$ minutes and $T_2=18$ minutes to complete a loop (excluding time stopped at bus stops), respectively. The first bus is the faster one, whilst the second bus is the slower one. We consider a busy period where $s=0.040$ people per second. A \emph{busy period} is defined by $k>\bar{k}$ in Eq.\ (\ref{kc}), where at least a pair of buses are persistently bunched. With these $T_1$ and $T_2$, we have the critical $k_c=\bar{k}=0.014$ ($\bar{k}=k_c$ since $N=2$). Note that we have included an overall factor of $1/2$ in Eq.\ (\ref{kc}) since alighting and boarding occur sequentially.

(Strictly speaking, more buses must be employed to meet the higher demand during busy times since each bus has a finite capacity, but we will ignore that limit for the purpose of investigating how a simple two-bus system performs during a busy period. The situation during a busy period with more buses is dealt with in Section 4.)
\item Frequency detuning in (b), during a lull period where $s=0.010$ people per second. A \emph{lull period} is defined by $k<\bar{k}$ in Eq.\ (\ref{kc}), where no buses are permanently bunched.
\end{enumerate}
Note that it suffices to consider one value of $s=0.010$ in (a) where the buses have identical natural frequency, since the behaviour of the bus system is the same for any fixed $s$. The different phases of lull and busy become distinct only when the system has frequency detuning \cite{Vee2019b}.

For each of the three situations 1, 2 and 3 as stated in the Introduction (Section\ \ref{sitofint}), we consider these setups (a), (b) and (c).

\subsection{No-boarding}\label{N=2}

The first situation is where buses are given the choices to \emph{stay} or \emph{leave} whenever they are at a bus stop, everybody who wishes to alight has done so, and there are people who would like to board. The reward $R_{NB}$ for each action (applicable to a system with any $N$ number of buses) is:
\begin{align}\label{RNB}
R_{NB}:=P + wf(\Delta\theta),
\end{align}
where $P$ is $1$ if a person is picked up and $0$ otherwise. Note that since the rate of people loading is $l=1$ person per second, either somebody boards or nobody boards at any time step of the simulation so this quantity is well-defined. The phase difference of the bus from the bus immediately behind it $\Delta\theta$ gives a reward defined by
\begin{align}\label{f}
f(\Delta\theta)=\left\{
\begin{array}{ll}
\frac{\Delta\theta}{360^\circ/N},\textrm{ if }\Delta\theta\leq\frac{360^\circ}{N}\\
1,\textrm{ otherwise}.
\end{array}
\right.
\end{align}
This function $f(\Delta\theta)$ which remains at $1$ beyond $360^\circ/N$ implies that the bus is doing fine and is not too slow, but is only receiving linearly diminishing reward if $\Delta\theta$ is smaller than $360^\circ/N$ which implies that it is too slow. The rationale for $f(\Delta\theta)$ staying flat instead of decreasing beyond $360^\circ/N$ is due to the fact that once there is nobody at the bus stop, it must leave, i.e. there is no option for it to lengthen its stay or try holding back. It is only when it is too slow ($\Delta\theta<360^\circ/N$) that it gets a lower reward.

A weight $w$ balances between $P$ (which encourages \emph{stay}) and $f(\Delta\theta)$ (which encourages \emph{leave} when $\Delta\theta<360^\circ/N$). Generally, a small $w\ll1$ leads to the buses eventually learning to behave like normal buses, where they would always $\emph{stay}$ since they always find somebody at the bus stop who wants to board. On the other hand, $w\gg1$ leads to the buses eventually learning to always \emph{leave} and maintain their perfectly staggered configuration of $\Delta\theta\approx360^\circ/N$. There is a finite range of $w\sim1$ where the bus system eventually learns to both pick up people and attain a reasonably staggered configuration such that the \emph{average waiting time} of the commuters at the bus stop is minimised. This precise range for $w$ depends on the particular conditions of the simulation environment like $s$, $l$, $T_i$, $N$, $M$. In each of our reinforcement learning runs, we set an appropriate $w$ in the balanced range. It appears that as long as $w$ is within this range, essentially identical qualitative results are obtained. In other words, the actual value of $w$ is unimportant as long as it is within that balanced range. (We will see later that the corresponding setup is not quite true for situation 2 on holding.)

\subsubsection{Identical natural frequency}

\begin{figure}
\centering
\includegraphics[width=13.5cm]{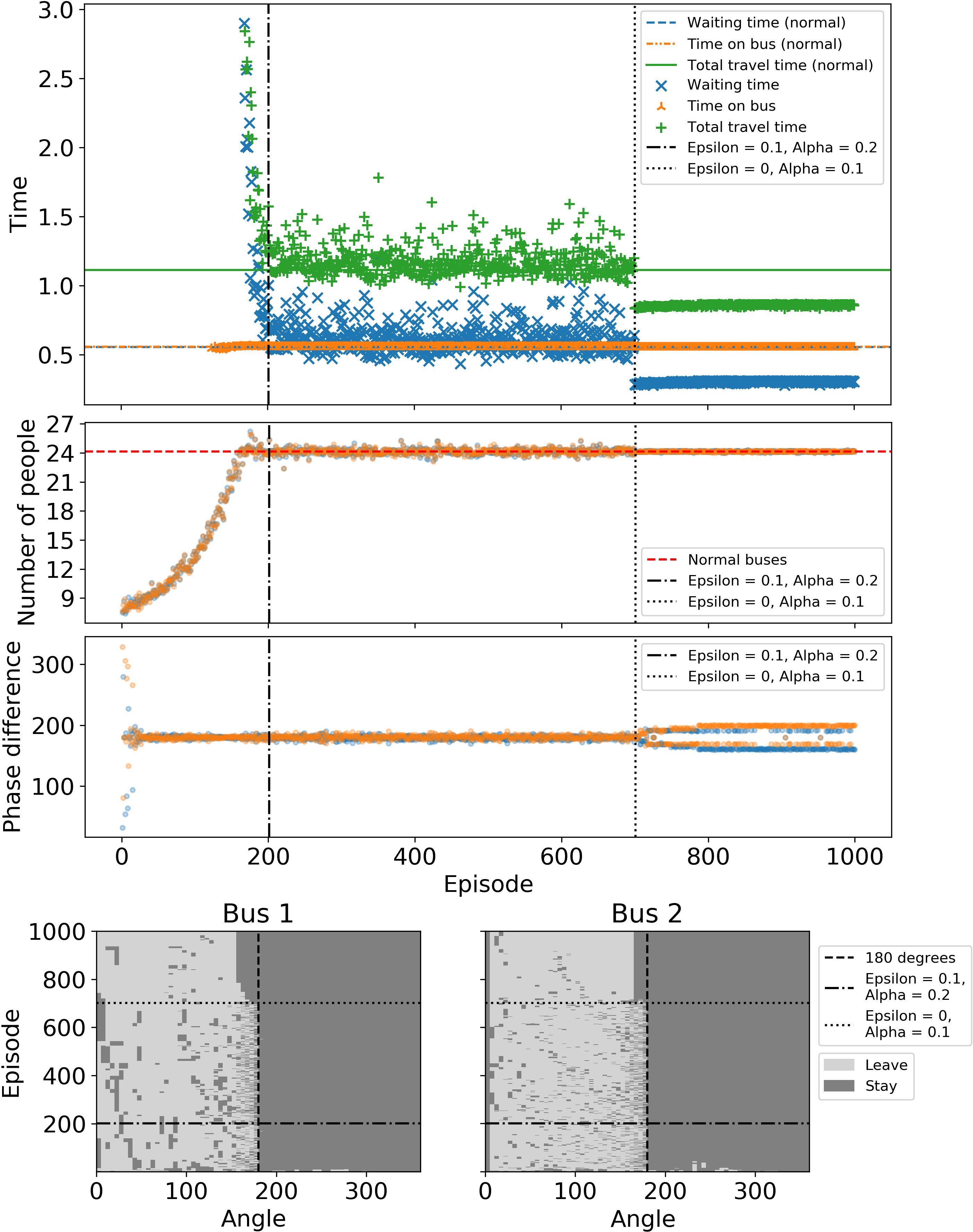}
\caption{Two buses with identical natural frequency serving a loop of bus stops learn the no-boarding strategy by reinforcement learning. Notably, they learn about the \emph{upper bound} at some angle \emph{strictly} less than $180^\circ$ to implement no-boarding, where if they exceed then they would inadvertently implement no-boarding too frequently such that they are not meeting the level of demand for service. Also, they learn to \emph{unbunch}, where one learns to \emph{stay} whilst the other learns to \emph{leave}, if their phase difference is $0^\circ$.}
\label{fig2}
\end{figure}

Fig.\ \ref{fig2} shows the results of these two buses with identical natural frequency undergoing reinforcement learning, comprising the average waiting time at the bus stop for a bus to arrive, average time spent on the bus, average total travel time (sum of the previous two quantities), and average number of people on the bus. In addition, it also shows the average phase difference from one bus as measured from the bus behind it, as well as the Q-tables for each bus.

The performance of the $N=2$ system is comparable to the analytical results in Ref.\ \cite{Vee2019b} where no-boarding is hard-coded, with an average waiting time of $\sim0.30$ units of $T$ during the last 300 episodes where the buses act greedily with respect to their learned Q-tables. Normal buses would typically end up bunching and the average waiting time is $\sim0.55$ units of $T$, so we see a nearly $50\%$ improvement.

Generally, a bus would implement no-boarding, i.e. \emph{leave} if $\Delta\theta<360^\circ/N$, and \emph{stay} otherwise. Remarkably, they also discover the following known result from Ref.\ \cite{Vee2019b}: There is an \emph{upper bound} on the angle to implement the no-boarding strategy, \emph{strictly} below the perfectly staggered angle of $360^\circ/N$. This upper bound arises because if the angle is too close to the staggered configuration, then the buses would end up implementing no-boarding too frequently at a rate where the people picked up is lower than the demand for service. The correspondence to the results from Ref.\ \cite{Vee2019b} is seen when the buses act greedily. On the other hand, in the $\varepsilon$-greedy phase before the $701$st episode, the upper bound is $\sim360^\circ/N$ because there is a probability of \emph{staying} which picks up people instead of stringently not boarding people. This is noted in the graph for the average phase difference and the Q-tables for the two buses: the buses are typically fluctuating around $\Delta\theta=180^\circ$ when exploration is involved, but this value is strictly less than $180^\circ$ in the fully exploitation phase. The performance before the $701$st episode is about those of normal buses with an average waiting time of $\sim0.55$ units of $T$ and sometimes even $1$ unit of $T$, instead of $\sim0.30$ units of $T$ in the last $300$ episodes.

Since the buses are randomly placed on the loop at the start of every episode, they may occasionally end up bunching. Can they unbunch? The answer is \emph{affirmative}. Since the buses are endowed with independent Q-tables, they learn \emph{opposite actions} if $\Delta\theta=0^\circ$: one bus \emph{stays} and the other \emph{leaves}. The system as a whole discovers a cooperative mechanism to correct itself when bunched. These results are consistently obtained in all five independent runs.

\subsubsection{Frequency detuning during busy period}

\begin{figure}
\centering
\includegraphics[width=13.5cm]{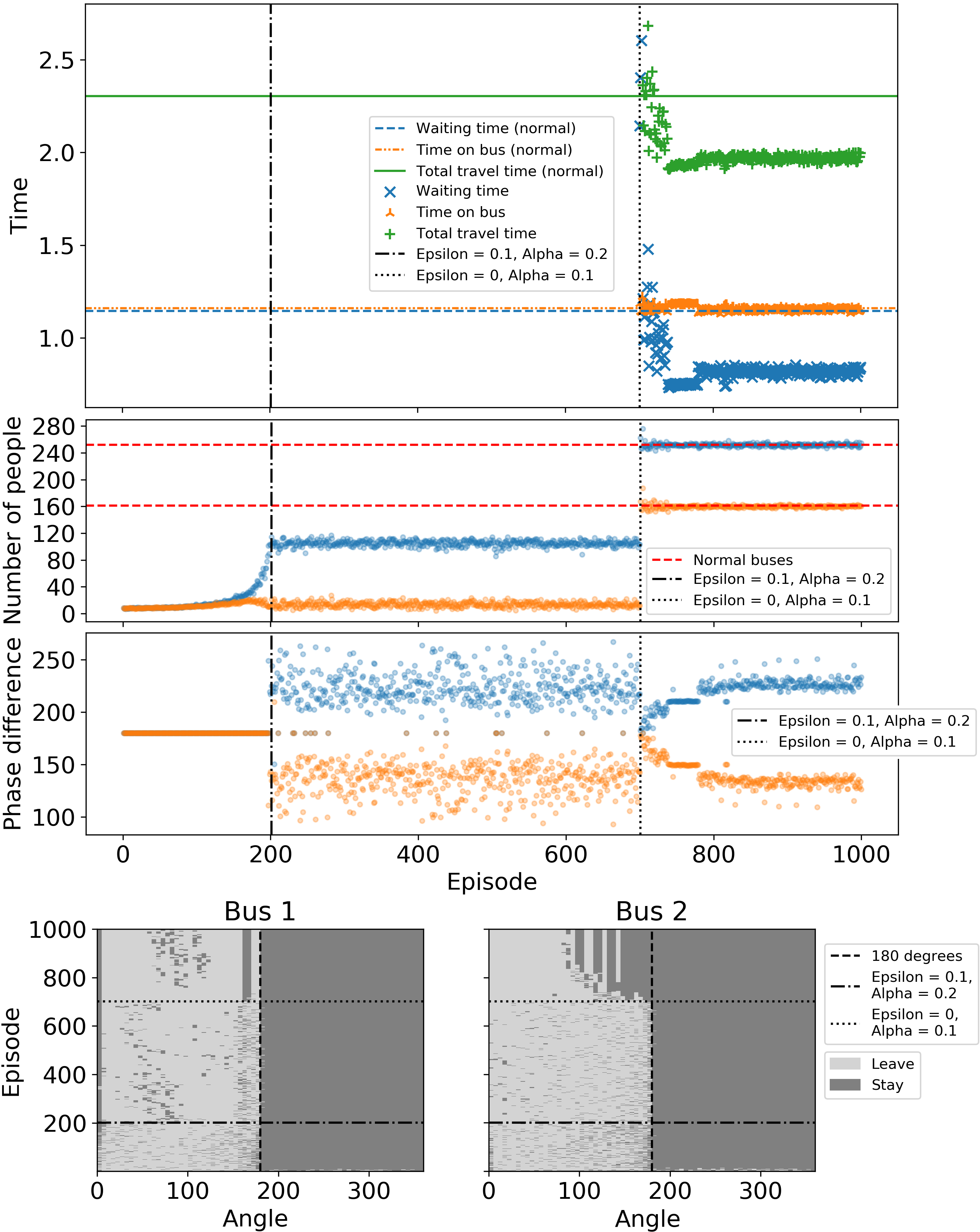}
\caption{Two buses with frequency detuning serving a loop of bus stops during a busy period learn the no-boarding strategy by reinforcement learning. The results are qualitatively similar to those in Fig.\ \ref{fig2}.}
\label{fig3}
\end{figure}

Fig.\ \ref{fig3} shows the corresponding results of these two buses with frequency detuning undergoing reinforcement learning during a busy period. The results here are essentially the same as the case with identical natural frequency, where the buses are able to learn the no-boarding policy. Bus 1 is the faster bus (in all frequency detuning cases for $N=2$, bus 1 is always the faster bus), and tends to pick up more people since the slower bus implements no-boarding and leave, leaving more people to the former to slow down its higher natural frequency. The average waiting time is also comparable to the results found in Ref.\ \cite{Vee2019b}. Incidentally, the absence of data points before the $701$st episode in the first graph is because the quantities are way too large due to the great number of people demanding service but not quite met by these two buses, such that they are beyond the range of the graph shown here.

Similar to the case with identical natural frequency, the two buses learn opposite actions when $\Delta\theta=0^\circ$ which would enable them to unbunch, and they also discover some upper bound strictly less than $180^\circ$ where no-boarding is implemented. The slower bus (bus 2) seems to find a lower value for the upper bound to implement no-boarding than the faster bus (bus 1), since it is the one which is usually slower and has to implement no-boarding. The slow bus should spend enough time at the bus stop to actually allow people to board, even though its phase difference $\Delta\theta$ may be less than $180^\circ$ otherwise it would not be picking up people and loses some reward for that. Therefore, the upper bound for it to implement no-boarding is lower to let this happen.

\subsubsection{Frequency detuning during lull period}\label{NBlull}

\begin{figure}
\centering
\includegraphics[width=13.5cm]{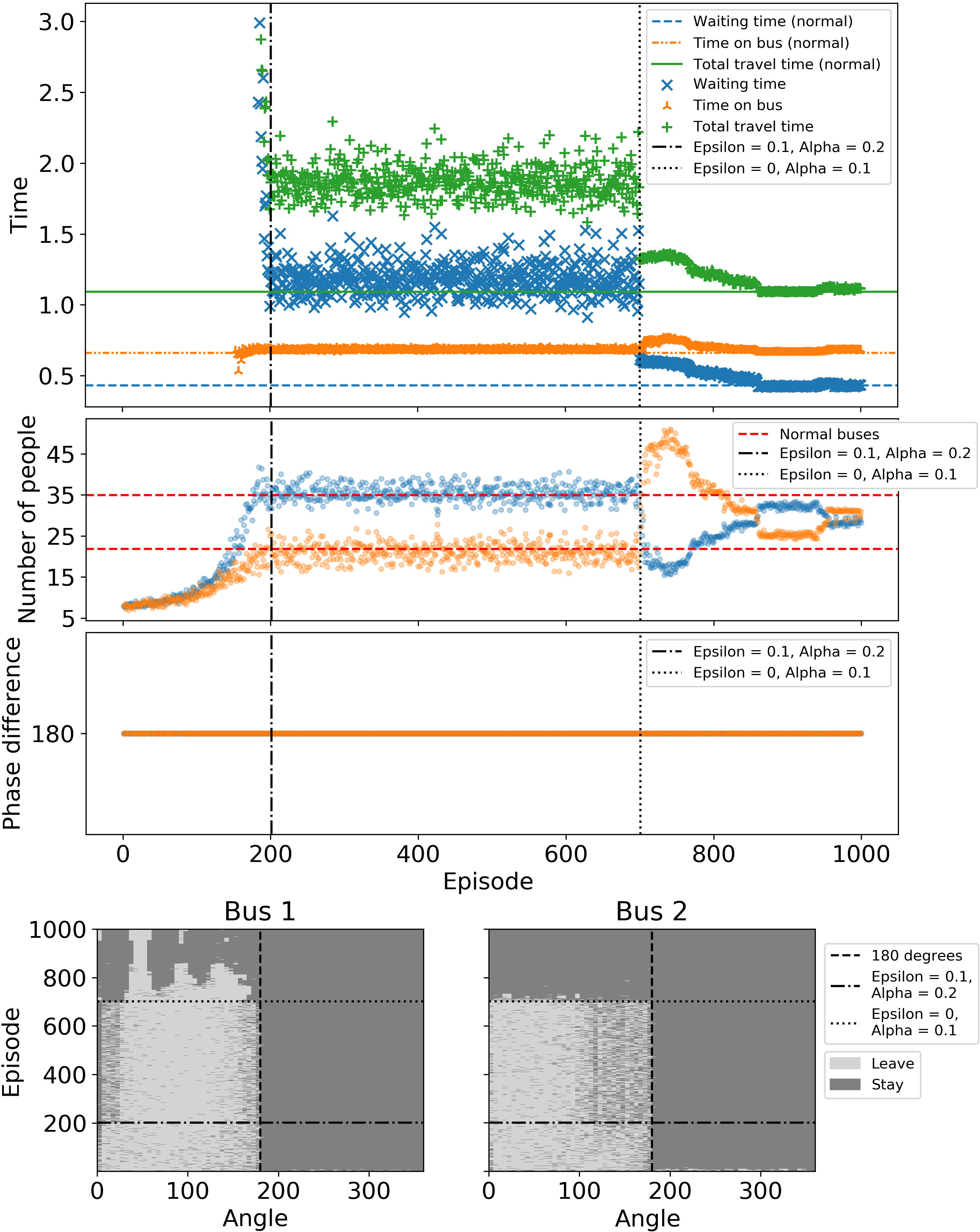}
\caption{Two buses with frequency detuning serving a loop of bus stops during a lull period \emph{does not learn the expected no-boarding strategy} by reinforcement learning.}
\label{fig4}
\end{figure}

Fig.\ \ref{fig4} shows the corresponding results of these two buses with frequency detuning undergoing reinforcement learning during a lull period. Here, the buses do not quite end up with the expected no-boarding strategy. This is in accordance to the observation noted in Ref.\ \cite{Vee2019b} where the no-boarding strategy backfires during the lull period because the slow bus has been sped up to the maximum by not picking up anybody! A hard-coded no-boarding policy would lead to the system effectively serving with one less bus because the slow bus almost always implements the no-boarding policy. Here, the buses found that perhaps it is better to just behave (almost) like normal buses, with performance that eventually matches closely to those of hard-coded normal buses. Incidentally, they do not necessarily need opposite actions when $\Delta\theta=0^\circ$ because their different natural frequencies allow them to unbunch.

Astonishingly, the optimal strategies for these two buses appear to defy what a human may intuitively conceive (at least initially), upon examining the Q tables of the buses (bottom plots in Fig.\ \ref{fig4}). When they begin to act greedily from the $701$th episode onwards (whilst still maintaining a learning rate of $\alpha=0.1$ so that they do continuously fine-tune their Q tables), the slow bus (bus 2) quickly changes to always behaving like a normal bus, with the \emph{fast bus (bus 1) implementing the no-boarding policy} when it is ``too slow''. Eventually by the $1000$th episode, it implements the no-boarding policy if $\Delta\theta\sim60^\circ$ from the bus behind it.

Perhaps the slow bus realises that there is no point for it to implement no-boarding as it simply cannot be sped up fast enough to overcome its lower relative velocity, and if it keeps leaving then it loses reward from not picking up people. On the other hand, the fast bus seems to think that when it is too slow, it should just \emph{leave} so that it can quickly attain $\Delta\theta\sim180^\circ$ which offers greater reward compared to getting stuck near $\Delta\theta\sim60^\circ$. If $\Delta\theta\ll60^\circ$, it probably would lose too much from not picking up people if it \emph{leaves}, before it can make $\Delta\theta$ grow up to $\sim180^\circ$ so that it would rather behave normally and just \emph{stay}; whilst if $\Delta\theta\gg60^\circ$, then the deficit in reward is not too high compared to $\Delta\theta\sim180^\circ$, such that it is fine with behaving normally and just \emph{stay} to earn the reward from picking up people. Hence in the lull period, we find that instead of trying in vain to keep the two buses staggered, they effectively \emph{increase the frequency detuning}.

With only the no-boarding strategy being studied in Ref.\ \cite{Vee2019b}, could the holding strategy or a holding + no-boarding strategy work to somehow provide some form of improvement for the bus system during the lull period? This is one primary aim of the framework in this paper, where we investigate reinforcement learning of the bus system to learn holding and holding + no-boarding strategies in the following subsections with $N=2$ buses serving a loop of bus stops.

\subsection{Holding}\label{holdingsubsection}

The second situation is where buses are given the choices to \emph{stay} or \emph{leave} whenever they are at a bus stop, everybody who wishes to alight has done so, and there is nobody at the bus stop. The reward $R_{H}$ for each action (applicable to a system with any $N$ number of buses) is:
\begin{align}\label{RH}
R_{H}:=g(\Delta\theta),
\end{align}
where
\begin{align}\label{g}
g(\Delta\theta)=\left\{
\begin{array}{ll}
\frac{1-\Delta\theta/360^\circ}{1-1/N},\textrm{ if }\Delta\theta>\frac{360^\circ}{N}\\
0,\textrm{ otherwise}.
\end{array}
\right.
\end{align}
This function $g(\Delta\theta)$ is discontinuous at $\Delta\theta=360^\circ/N$. It is $0$ at and less than $360^\circ/N$ since it is regarded as ``slow'' with respect to the bus behind it. On the other hand, it approaches $1$ from the right if $\Delta\theta>360^\circ/N$ since this is the ideal phase difference that it should strive for when it is ``too fast''. From a reward of $1$ just over $360^\circ/N$, it then linearly decreases to $0$ as $\Delta\theta$ grows towards $360^\circ$ since larger phase difference is getting away from ideal.

This reward $R_H$ does not say anything about how long it will repeatedly \emph{stay} at a bus stop. One option is to include a negative reward so that buses do not simply remain at a bus stop indefinitely, i.e. $-1$ for each \emph{stay} action or if a certain number of consecutive \emph{stay} actions are executed (recall that in this situation, everyone who wishes to alight has done so, and there is nobody at the bus stop hence a negative reward discourages ``time wasting''). Then, a weight $w_H$ could be introduced between this negative reward and $g(\Delta\theta)$, analogous to the no-boarding reward $R_{NB}$ in Eq.\ (\ref{RNB}). The hope with this is that there is some balanced region for $w_H$ such that the bus system does not excessively remain at a bus stop. It turns out that Eq.\ (\ref{RH}) works well and a bus does not indefinitely remain at a bus stop because it will get $0$ reward if $\Delta\theta\leq360^\circ/N$ which would prompt it to \emph{leave}. Furthermore, unlike the no-boarding case where the actual value of $w$ does not change the outcome as long as $w$ is in the balanced region, here different values of $w_H$ would lead to different durations a bus may hold at a bus stop. We find this to be equivalent to just imposing a limit on how long a bus can hold.

Note also that this situation is different from the no-boarding situation in the following sense: In the no-boarding situation, if a bus chooses the ``unconventional'' action of \emph{leave} when there is somebody to pick up, then that is the end for this round at this current bus stop. It leaves the bus stop and moves on. However for the holding situation here, if a bus chooses the ``unconventional'' action of \emph{stay} when there is nobody to pick up, then it gets to choose its action again at this current bus stop. This is why the nature of the rewards as well as their emergent behaviours (as we will see below for holding) are not directly analogous.

\subsubsection{Identical natural frequency}

\begin{figure}
\centering
\includegraphics[width=13.5cm]{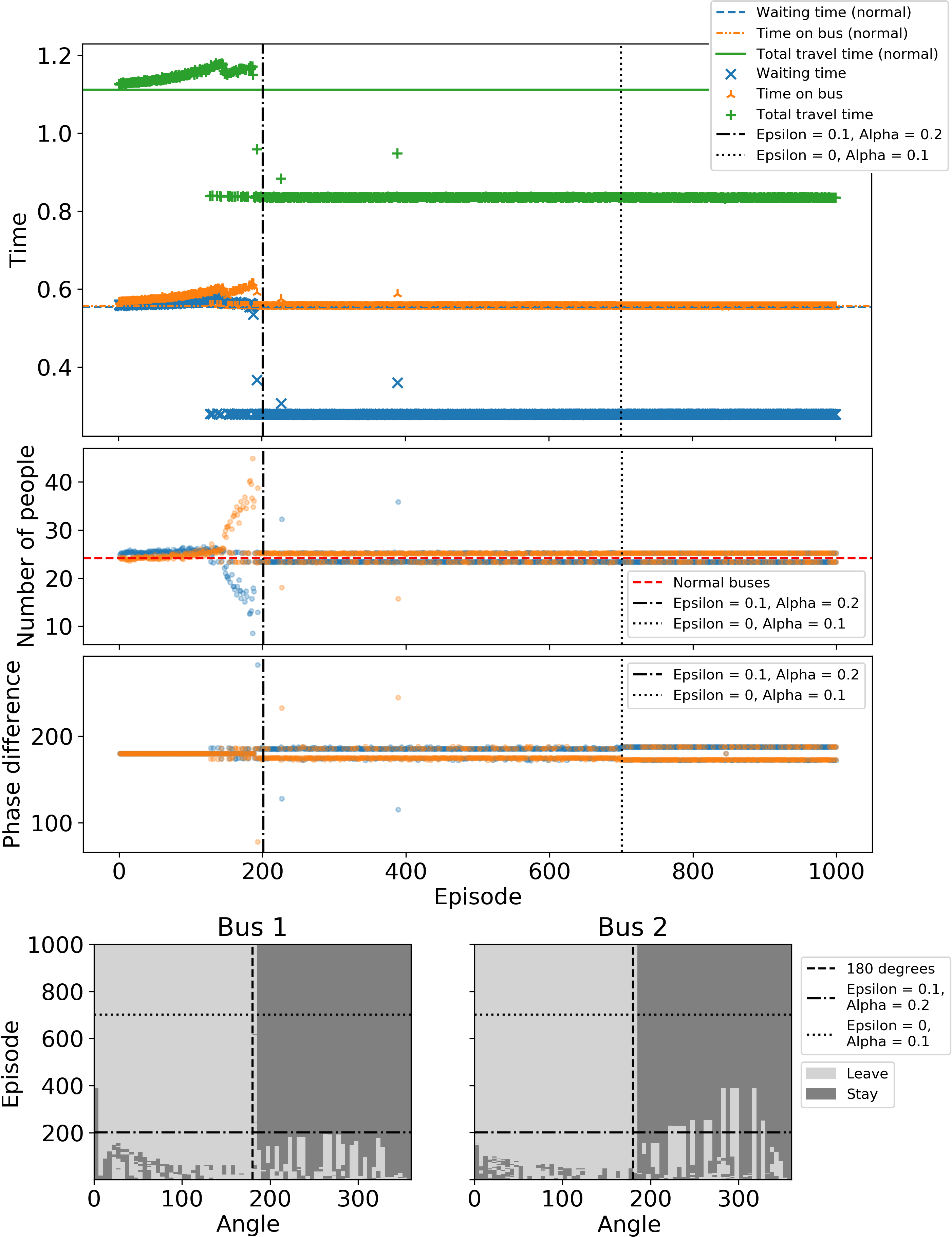}
\caption{Two buses with identical natural frequency serving a loop of bus stops learn the holding strategy by reinforcement learning.}
\label{fig5}
\end{figure}

Fig.\ \ref{fig5} shows the results of these two buses with identical natural frequency undergoing reinforcement learning, corresponding to the graphs in the previous figures. The holding strategy looks impressively effective, where it achieves sub-$0.3$ units of $T$ for the average waiting time of the commuters, even before the $201$st episode where $\varepsilon$ decays to $0.1$ and $\alpha=0.2$. The way the holding strategy works is that if $\Delta\theta$ is not reasonably close to $360^\circ/N$, then the faster bus would remain at the bus stop until $\Delta\theta\sim360^\circ/N$. Since there is not much noise in the simulation environment (no traffic, rate of people arrival at bus stops is constant), the two buses would just remain fairly staggered thereafter.

The buses learn that there is a \emph{lower bound} to implement the holding strategy, which is strictly larger than $360^\circ/N$. This is the consequence of the discontinuity at $360^\circ/N$ in the reward $R_H$ in Eq.\ (\ref{RH}) where its value at $360^\circ/N$ itself is 0, which discourages \emph{staying}. The buses also learn to never \emph{stay} for any phase difference $\Delta\theta\leq360^\circ/N$ as that gives $0$ reward. This is important because the two buses may be exactly staggered with $\Delta\theta=180^\circ$ and if they both learn to stay, then they would just stay forever.

Curiously, the buses ostensibly learn opposite actions when they bunch, i.e. $\Delta\theta=0^\circ$, during some earlier episodes but these opposite actions disappear in subsequent episodes and both end up learning to \emph{leave} if $\Delta\theta=0^\circ$. How do they unbunch then, if they both take the same action? Since their positions on the loop are randomised at the start of each episode, they would inevitably end up bunching at some point. Upon closer inspection, there is actually a natural mechanism for normal buses to \emph{momentarily} unbunch: If there is an odd number of people at the bus stop (or the number of people modulo $N$ is not zero, for $N$ bunched buses in general), then at that instant one bus stays there to pick up that last person whilst the other bus sees nobody and leaves. (Note: In the simulation, a loop over all buses is carried out at that instant. The first bus sees somebody and stays to pick up, whilst the second bus then sees nobody and leaves.) This is how normal buses can momentarily unbunch from $\Delta\theta=0^\circ$ to be $\Delta\theta=5^\circ>0^\circ$. (Recall that we discretise the angles by $72$ bins of $5^\circ$.) Of course, at the next bus stop they would swiftly bunch again since they have to allow people to alight. So normal buses remain bunched. Even buses implementing the no-boarding strategy would remain bunched, which is why they learn opposite strategies via the process of reinforcement learning over the $1000$ episodes.

However, the holding strategy differs in this unique manner: When a pair of bunched buses naturally unbunch due to one bus staying to pick up somebody and the other bus leaving as it has nobody to pick up, then the bus that stayed would see its phase difference as measured from the bus behind it (which is the bus that has just left, in front of it) to be $\Delta\theta=355^\circ$, implying that it is \emph{way too fast}! Therefore, it would implement \emph{stay} all the way based on its Q-table, until $\Delta\theta=185^\circ$. This is why there is no need for these buses implementing the holding strategy to have to learn opposite strategies.

\subsubsection{Frequency detuning during busy period}\label{holdingbusy}

\begin{figure}[ht]
\centering
\includegraphics[width=13.5cm]{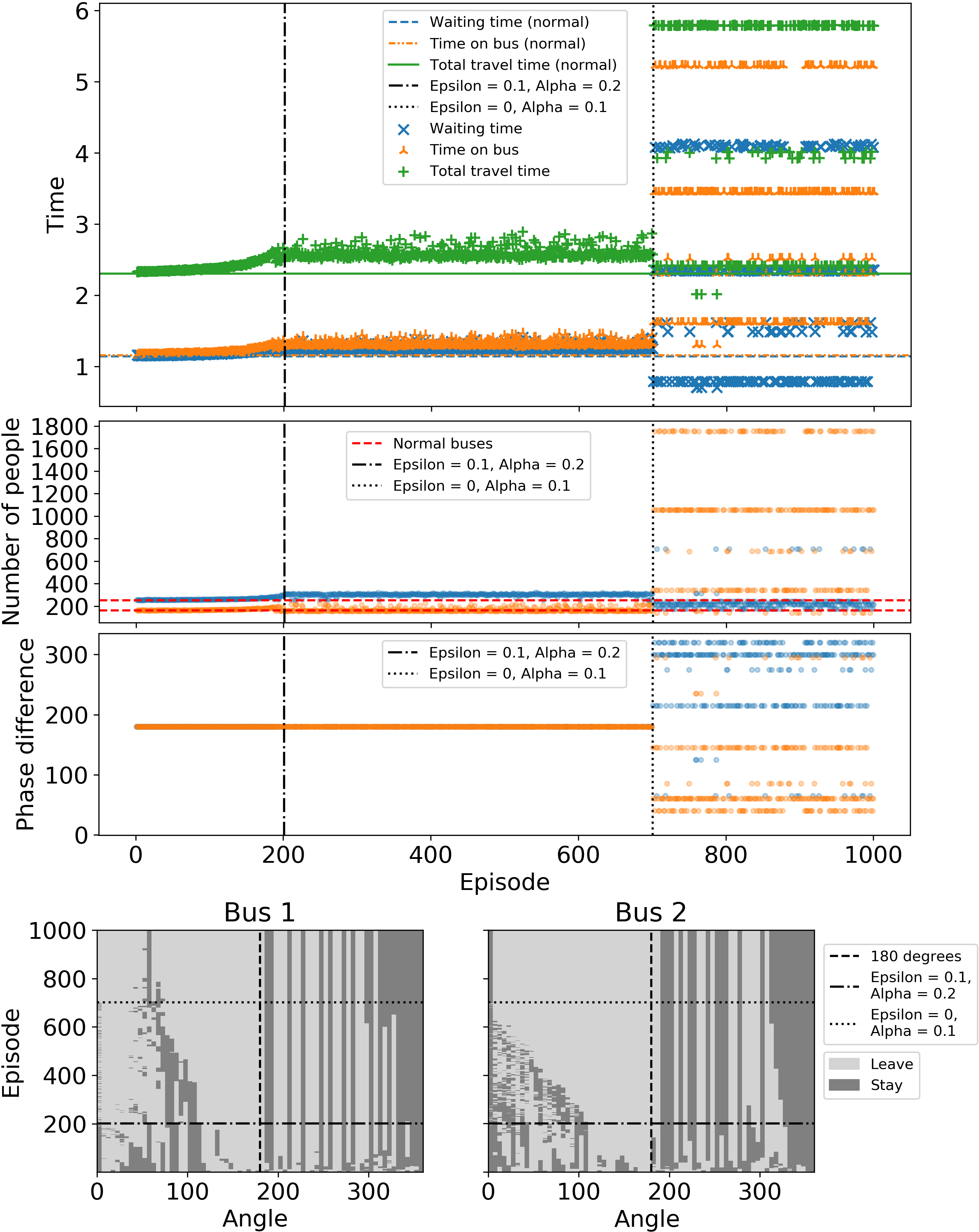}
\caption{Two buses with frequency detuning serving a loop of bus stops during a busy period learn the holding strategy by reinforcement learning. The entire system is slowed down greatly and generally performs worse than normal buses.}
\label{fig6}
\end{figure}

Fig.\ \ref{fig6} shows the results of these two buses with frequency detuning undergoing reinforcement learning, corresponding to the graphs in the previous figures. Here, demand for service is high in a busy period. The holding strategy slows down the fast bus, effectively slowing down the entire bus system. Since the reward for the system is purely to keep $\Delta\theta$ staggered, it does not care about staying too long at bus stops and the average waiting time suffers. This is indicated by the average number of people on the bus blowing up into thousands! Since there are only $N=2$ buses trying to meet a high demand during the busy period, mistakes made when the buses explore other actions would lead to many of the other $M=12$ bus stops rapidly accumulating people waiting for service. We will see in Section \ref{manybuses} that with $N=6$ buses in the busy period, there are sufficient buses going around such that they are able to reasonably learn the holding strategy. With more buses, mistakes made by one bus when it explores is covered by other buses such that the number of people waiting at the $M=12$ bus stops do not blow up.

The overall performance here is generally worse than normal buses. The average time that commuters spend on the bus also suffers since the buses expend more time at each bus stop before they get off at their respective destinations. This is one drawback of the holding strategy where the system gets slowed down, which is why a no-boarding strategy is arguably superior in a busy period.

Oddly enough for the holding strategy, this time in the busy period, the buses learn opposite actions when they bunch with $\Delta\theta=0^\circ$. Here, they have to learn to unbunch deliberately because the busy period would otherwise keep them persistently bunched.

\subsubsection{Frequency detuning during lull period}\label{holdinglull}

\begin{figure}[ht]
\centering
\includegraphics[width=12cm]{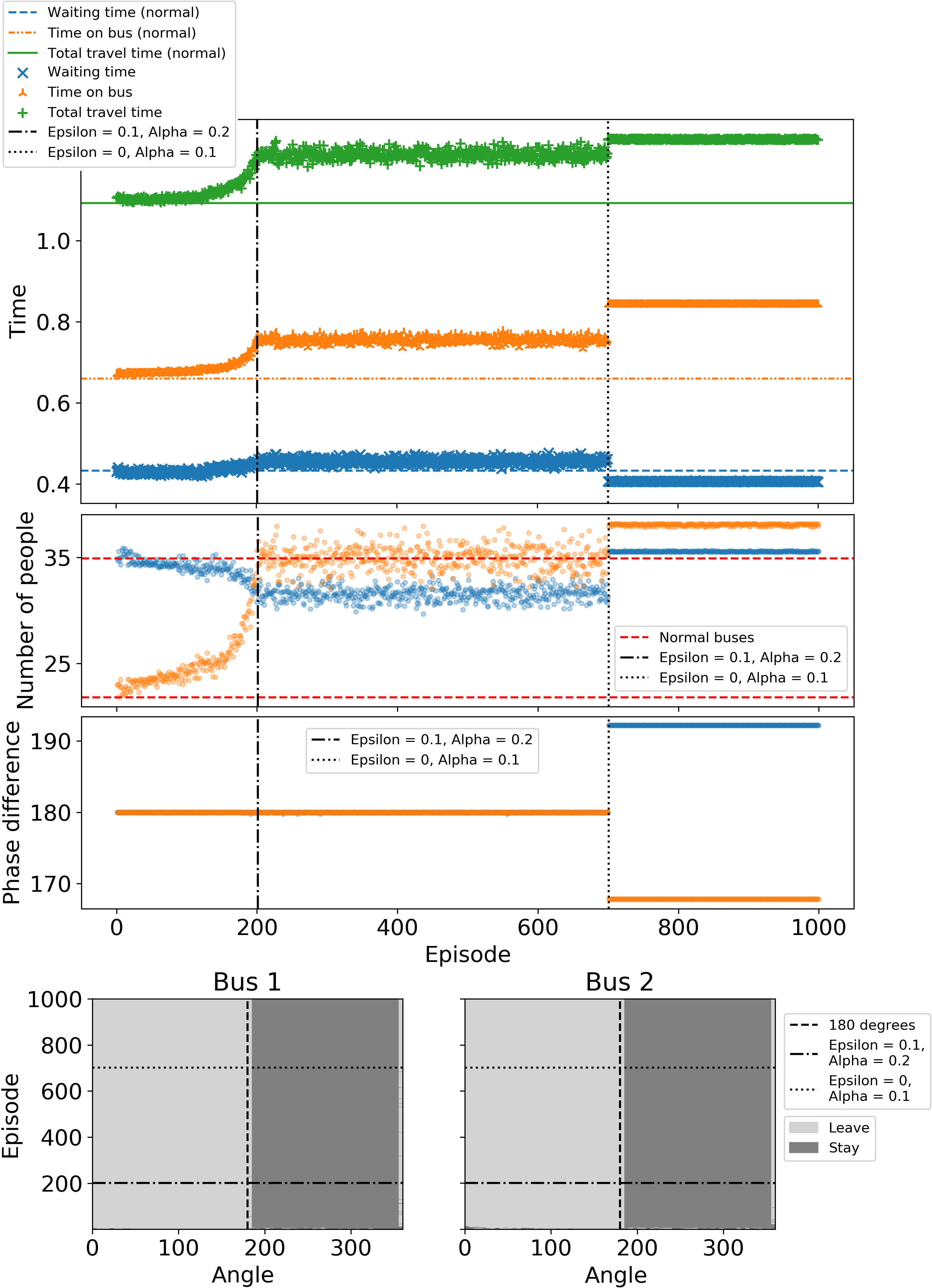}
\caption{Two buses with frequency detuning serving a loop of bus stops during a lull period learn the holding strategy by reinforcement learning. It is able to slightly improve the average waiting time, at the expense of slowing down the fast bus such that the average time spent on bus and average total travel time are increased.}
\label{fig7}
\end{figure}

\begin{table}
\centering
\begin{tabular}{|c|c|c|}
\hline
Strategy & No-boarding & Holding\\
\hline
Identical frequency & Positive & Positive\\
Frequency detuning, busy & Positive & Negative, Q-tables unable to completely train\\
Frequency detuning, lull & Negative & Improves waiting time, but adds time spent on bus\\
\hline
\end{tabular}
\caption{Qualitative performance of the no-boarding and holding strategies, respectively, in various setups of $N=2$ buses serving $M=12$ bus stops in a loop.}\label{table1}
\end{table}

Fig.\ \ref{fig7} shows the results of these two buses with frequency detuning undergoing reinforcement learning, corresponding to the graphs in the previous figures. For the first time in a lull period, we find a way to improve the average waiting time of commuters, by means of a holding strategy. However, the cost involved is that commuters would spend more time on the bus, on average, since the way the holding strategy works in keeping the buses staggered is by delaying the fast bus to the extent of being as slow as the slow bus. The average number of people on the fast bus is closer to that on the slow bus when the holding strategy is implemented, as compared to the normal buses where one bus consistently picks up more people than the other.

In spite of increasing the average time spent on bus and the average total travel time, perhaps the holding strategy may be viewed as viable since it is arguably less of a pain point to be on the bus enjoying the air conditioner compared to being out at the open bus stop where it may be hot under the blazing sun, wet during a thunderstorm, or even chilly during winter (in countries with four seasons).

We summarise the qualitative performance of the no-boarding and holding strategies for each of the setups that we have discussed in Table\ \ref{table1}. Quantitative percentage improvement (or worsening) will be given in the next section where we consider a system with $N=6$ buses serving $M=12$ bus stops in a loop.

\subsection{Combined no-boarding and holding strategies}\label{2combined}

The third situation is where buses are given the choices to \emph{stay} or \emph{leave} whenever they are at a bus stop and everybody who wishes to alight has done so. Here, ``somebody wants to board'' and ``nobody wants to board'' are distinct. Therefore, we take these situations as a combination of the first two situations where situation 1 occurs when there is somebody who wants to board, and situation 2 occurs when nobody wants to board.

\begin{figure}
\centering
\includegraphics[width=13.5cm]{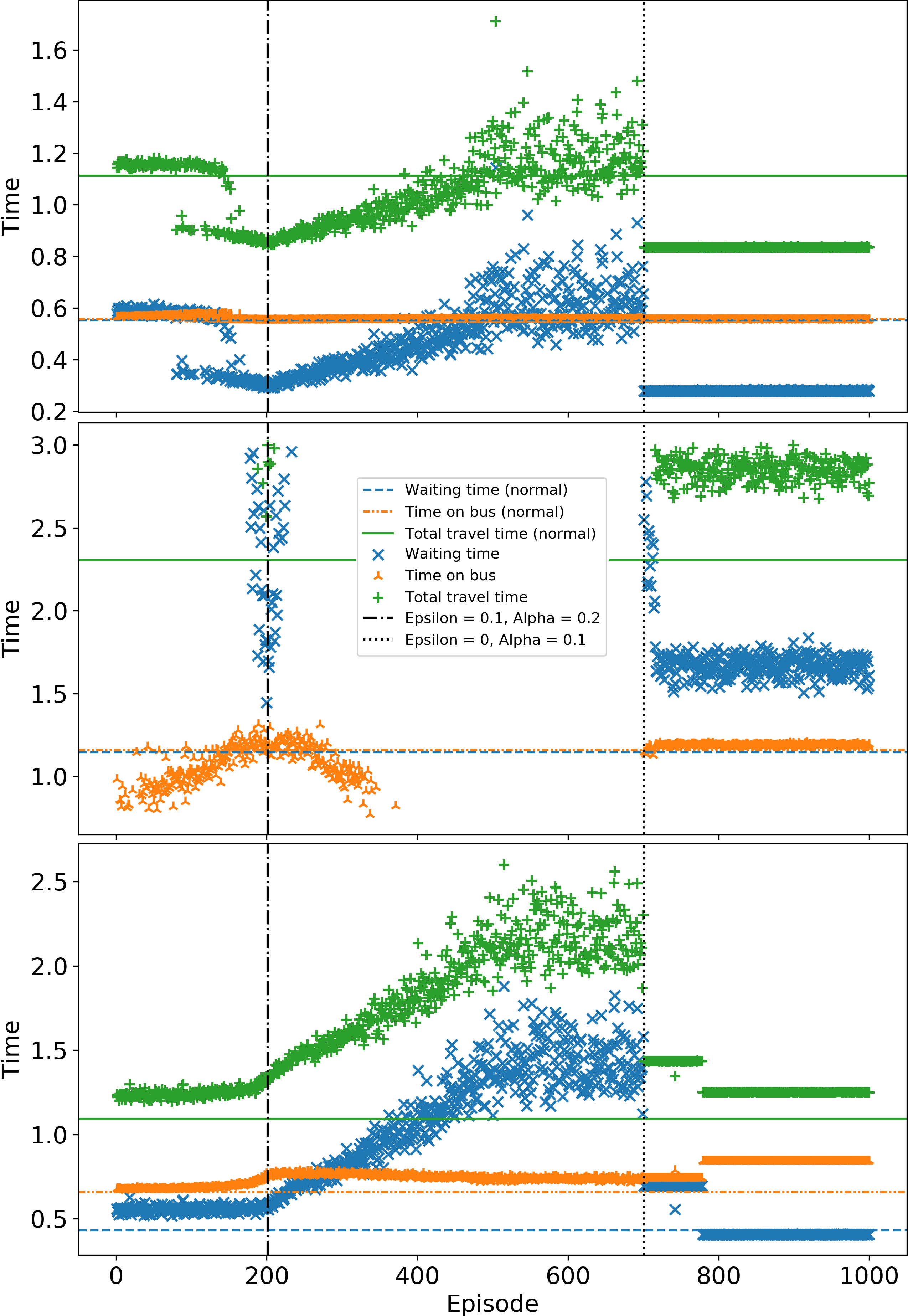}
\caption{$N=2$ buses serving $M=12$ bus stops in a loop, where a bus takes an action at a bus stop when nobody wants to alight. Top: identical natural frequency, middle: frequency detuning during busy period, bottom: frequency detuning during lull period. All three graphs share the same legend.}
\label{fig8}
\end{figure}

Note that since situation 2 can only occur after everybody at the bus stop has been picked up, if the bus leaves when somebody is still there, then that is the end for this round at the bus stop and situation 2 is completely sidestepped. After 1000 episodes of training, we find that the buses' Q tables for situation 1 is trained but those for situation 2 are not. To allow for a fair amount of training for the latter Q tables, we implement the following additional exploration possibility: In the first 200 episodes, if a bus chooses to \emph{leave} when there is somebody to pick up, then there is a probability of $\Upsilon=0.9$ that it switches to \emph{stay}. From the $201$st to $500$th episode, $\Upsilon$ is linearly decayed from $0.9$ to $0$. A reasonably high value of $\Upsilon$ is necessary to expose the buses to situation 2 for training, because for example if there are $10$ people at the bus stop, then a bus needs to choose $10$ consecutive \emph{stay} actions before it has the chance to encounter and train for situation 2.

Fig.\ \ref{fig8} summarises the results for the three setups (a), (b) and (c) listed right at the beginning of this section (before the start of Section \ref{N=2}) corresponding to having identical natural frequency, frequency detuning in the busy as well as in the lull periods, respectively. We find that the buses are indeed able to learn both Q tables such that each Q table resembles that in the corresponding situations in Sections \ref{N=2} and \ref{holdingsubsection}, with some minor differences that account for the fact that the buses can decide on \emph{stay} or \emph{leave} in two different situations 1 and 2. In terms of the performance, the graphs near episode $\sim200$ are primarily dominated by the holding strategy (recall that here, the performance is as good as fully exploiting even though $\varepsilon=0.1$), whilst those approaching episode $\sim500$ are primarily dominated by the no-boarding strategy (recall that here, the performance is not as good as fully exploiting since $\varepsilon=0.1$ induces the bus to \emph{leave} when it should not, leaving the people behind to unnecessarily wait for the next bus). The performance transitions between that of holding to no-boarding somewhere between episodes $200$ to $500$ as $\Upsilon$ decays from $0.9$ to $0$. Finally in the last $300$ episodes where there is no longer any exploration $\varepsilon=0$, the bus system settles into their most optimal strategy that they have acquired from the possible combinations.

In the case with identical natural frequency, the buses \emph{harness both the no-boarding and holding strategies} where a bus would implement no-boarding if it is ``too slow'' ($\Delta\theta<360^\circ/N$) and implement holding if it is ``too fast'' ($\Delta\theta>360^\circ/N$). For the busy period with frequency detuning, however, since the buses are not able to train the Q tables corresponding to holding, it does not actually execute the holding strategy properly. This results in the extended waiting times, similar to what happened with the holding strategy alone during a busy period in Section\ \ref{holdingbusy}.

For the two buses with frequency detuning during the lull period, it turns out that the buses perform as good as the holding strategy in terms of improving the average waiting time of commuters at the bus stop for a bus to arrive. The buses are able to learn that the no-boarding strategy, when applied by the slow bus (bus 2) in an attempt to speed it up, would result in it not picking up sufficient passengers and nullify its whole purpose of serving the loop. Since the frequency detuning is too large compared to the demand level, the no-boarding policy alone cannot speed it up, as we have seen in Section\ \ref{NBlull}. Here with both the options to \emph{stay} and \emph{leave} when there is somebody as well as nobody who wants to board, the buses try to harness both no-boarding and holding strategies, such that a bus would implement no-boarding if it is ``too slow'' ($\Delta\theta<360^\circ/N$) and implement holding if it is ``too fast'' ($\Delta\theta>360^\circ/N$). However, the buses realise that the combination of no-boarding and holding is not the most optimal way to go, since bus 2 would be leaving with few people when it implements no-boarding and relatively little gain in speeding up. Eventually somewhere close to the $800$th episode, the slow bus decides that it is better to forget about no-boarding even if $\Delta\theta<360^\circ/N$, and the bus system relies entirely on the fast bus (bus 1) to implement the holding strategy. The performance then matches with the purely holding strategy presented in Section\ \ref{holdinglull}, with identical improvement in average waiting time, and identical increases in average time spent on bus as well as average total travel time.

In summary, given both possibilities of situations 1 and 2, the bus system is able to find the revelant most optimal strategy depending on the particular conditions. For example with identical natural frequency, they harness both the no-boarding and holding strategies to the fullest. On the other hand with frequency detuning in the lull period, they revert to the holding strategy and ditch the no-boarding policy.

\section{Any \texorpdfstring{$N$}{N} buses serving \texorpdfstring{$M$}{M} bus stops in a loop}\label{manybuses}

This framework is directly generalisable to any $N$ buses serving $M$ bus stops in a loop. We have carried out more simulations with $N=3$ buses and even $N=6$ buses, respectively. System with many buses generally produce qualitatively similar results to those already discussed for the case with $N=2$ buses.

Table \ref{table2} summarises the quantitative performance (in terms of average waiting time for a bus to arrive at a bus stop) for these $N=6$ buses serving $M=12$ bus stops under various conditions and setups. The setups with identical natural frequency is given $k=0.010$, busy with frequency detuning is given $k=0.063$, and lull with frequency detuning is given $k=0.010$. When there is frequency detuning, the six buses are prescribed different natural frequencies, selected within the range of 12 minutes to 18 minutes (excluding time stopped at bus stops).

\begin{table}
\centering
\begin{tabular}{|c|c|c|}
\hline
No-boarding & Average waiting time ($T=12$ minutes) & Percentage change \\
\hline
Identical frequency & $0.511$ (normal) to $0.180$ & Improves by $64.8\%$\\
Frequency detuning, busy & $0.793$ (normal) to $0.340$ & Improves by $57.1\%$\\
Frequency detuning, lull & $0.174$ (normal) to $0.290$ & Aggravates by $66.7\%$\\
\hline
\end{tabular}
\vskip0.5cm
\begin{tabular}{|c|c|c|}
\hline
Holding & Average waiting time ($T=12$ minutes) & Percentage change \\
\hline
Identical frequency & $0.511$ (normal) to $0.086$ & Improves by $83.2\%$\\
Frequency detuning, busy & $0.793$ (normal) to $0.160$ & Improves by $79.8\%$\\
Frequency detuning, lull & $0.174$ (normal) to $0.126$ & Improves by $27.6\%$\\
\hline
\end{tabular}
\vskip0.5cm
\begin{tabular}{|c|c|c|}
\hline
Combined & Average waiting time ($T=12$ minutes) & Percentage change \\
\hline
Identical frequency & $0.511$ (normal) to $0.180$ & Improves by $64.8\%$\\
Frequency detuning, busy & $0.793$ (normal) to $0.300$ & Improves by $62.2\%$\\
Frequency detuning, lull & $0.174$ (normal) to $0.126$ & Improves by $27.6\%$\\
\hline
\end{tabular}
\caption{Quantitative performance of the no-boarding strategy, holding strategy, and a combination of them, respectively, in various setups of $N=6$ buses serving $M=12$ bus stops in a loop.}\label{table2}
\end{table}

Here are some noteworthy points in the \emph{lull period} where the buses have \emph{frequency detuning}:

\begin{enumerate}
\item Holding positively improves the average waiting time by $27.6\%$ whereas no-boarding backfires and lengthens by $66.7\%$. Nevertheless, holding would increase the average time spent on bus by $24.4\%$ (from $0.624T$ to $0.776T$) such that the average total travel time would increase by $13.0\%$ (from $0.798T$ to $0.902T$).

\item A combination of no-boarding and holding positively improves the average waiting time by $27.6\%$, matching the savings due to pure holding. Nevertheless, this combination would increase the average time spent on bus by $20.4\%$ (from $0.624T$ to $0.751T$) such that the average total travel time would increase by $9.9\%$ (from $0.798T$ to $0.877T$).
\end{enumerate}

With more buses, the ability to keep buses staggered over the loop significantly divides off the average waiting time at the bus stop for a bus to arrive. The expense incurred with the increase in average time spent on bus and also the average total travel time by implementing the holding strategy during the lull period seems to be well worth it, when described in terms of percentages. Apart from that, the combined strategies saves as much time as holding on the average waiting time whilst incurring \emph{less cost on average time spent on bus (and average total travel time)}. Hence, we see a further improvement thanks to a combination of no-boarding and holding strategies during the lull period for this system with $N=6$ buses.

\section{Discussion and concluding remarks}

The use of reinforcement learning for a bus system serving a loop of bus stops has shown the potential of discovering strategies to optimise performance of the system. The framework employed in this paper takes advantage of the phase difference between buses in a loop where maintaining a staggered configuration translates to minimising the average waiting time of commuters at the bus stops for a bus to arrive. This provides a way to deal with the high variance of the individual waiting times that causes convergence of the Q-table to be essentially impossible within reasonable simulation time (or perhaps not even possible in some cases \cite{Sato02}).

The system has learnt that no-boarding and holding strategies are indeed both useful strategies when the buses have identical natural frequency. No-boarding speeds up the slower bus whilst holding slows down the faster bus. The former is also useful when buses have frequency detuning during the busy period but the latter may slow down the system too much. Nevertheless, the holding strategy is salutary in the lull period where the fast bus is slowed down to match the slow bus in order to maintain a reasonably staggered configuration, at the expense of increasing time spent on bus and total travel time. This offers a solution during the lull period, where the no-boarding strategy simply does not work at all.

It is interesting to note that although the buses are given \emph{low level actions} of \emph{stay} or \emph{leave} at the bus stop when nobody wants to alight --- essentially \emph{knowing only the rules of the game}, reinforcement learning leads to the discovery of \emph{high level strategies} of no-boarding \cite{Vee2019b,Vee2019c} and holding \cite{Abk84,Ros98,Eber01,Hick01,Fu02,Bin06,Daganzo09,Cor10,Cats11,Gers11,Bart12,Chen15,Moreira16,Wang18}. This illustrates the utility of a reinforcement learning framework where the system is able to arrive at high level strategies without human presumptions and priors, like how the \emph{AlphaZero} programme \cite{AlphaZero} is able to come up with and even validate known human strategies and tactics (e.g. the Berlin defence against the Ruy Lopez in Chess), discrediting some of them (e.g. the French defence in Chess, apparently) and even revealing new possibilities (e.g. sacrificing multiple pawns and pieces in favour of long-term subtle activity in Chess --- highly impressing many Chess Grandmasters, including a former World Champion \cite{Kasparov,GC}).

In particular, these intelligent buses are able to behave cooperatively to unbunch in unique and interesting ways like learning opposite actions in the case of no-boarding, as well as one bus just holding to allow the other to correct their phase difference. They also discover useful strategies with the appropriate bounds where no-boarding and holding are implemented. These emergent behaviours arise from the ability of the buses to learn and improve from their interactions, eventually settling into some collectively optimal strategies. On top of that, the system also makes use of combining the options appropriately in various setups. This is important when we move on to non-stationary environments where the system \emph{must} encounter various situations and be able to act with an optimal strategy.

Being low level however, implies that the bus system does not actually ``know'' that it can ``choose to implement a no-boarding strategy or a holding strategy'' at will. All that it cares is: A bus is at a bus stop, nobody wants to alight. Is there anybody who wants to board? If yes, then should it \emph{stay} or \emph{leave}? If not, then should it \emph{stay} or \emph{leave}? Since it typically encounters somebody who wants to board, if it \emph{leaves}, then it will not encounter the latter situation where nobody wants to board --- thus sidestepping the holding option. In order to allow for a balanced combination between a no-boarding strategy and a holding strategy, we have augmented the exploration phase with a new hyperparameter $\Upsilon$. Alternatively, perhaps a different approach with this framework to be \emph{higher level} would allow for faster convergence of the Q table. In other words, when a bus is at a bus stop, it is ``conscious'' about the options on: (a) behaving like a normal bus; (b) \emph{leave} --- implement no-boarding; or (c) \emph{stay longer} --- implement holding. This therefore places the options for no-boarding and holding on an equal footing, alongside behaving normally, and alleviates the bias towards implementing no-boarding over holding. Nevertheless, we have shown here that these low level actions do lead to the high level no-boarding and holding strategies, in the situations where somebody wants to board and nobody wants to board, respectively. This establishes the mechanisms on how low level actions lead to the emergence of high level coordinated strategies of the buses. With higher level actions, the faster convergence becomes a crucial utility for being adaptive in non-stationary environments of the real world.

Thus far, this paper assumes that all $M=12$ bus stops are perfectly staggered around the loop and all have the same rate of people arrival, $s$. We also imposed that each person wants to head to an antipodal destination. Whilst seemingly simplified, this represents an important first step in a series of increasingly complex progression for our research on the bus system undergoing reinforcement learning. In particular, we have established and clarified the behaviour of the bus system with identical natural frequency, as well as with frequency detuning in the busy and lull periods. Each setup has distinct characteristics of its own and the appropriate strategy should be applied especially if there is frequency detuning, viz. no-boarding during busy and holding during lull.

A step forward would be to generalise the environment based on real data that we have collected in Ref.\ \cite{Vee2019}, to investigate how the bus system may arrive at novel and even adaptive strategies to deal with non-stationary environments where people may wish to head towards some \emph{hubs} at certain times of the day, with some bus stops having higher rates of people arrival, i.e. $s_i$ for each $i=1,2,\cdots,M$. The framework in this paper serves as a good platform for greater layers of complexity to be piled up on the environment. Eventually, we could then implement such strategies to our Nanyang Technological University campus shuttle bus service upon where this environment is modelled after \cite{Vee2019,NTUnews,NTUautobuses}, and subsequently even adapt to more complex bus routes.

\begin{acknowledgments}
This work was supported by the Joint WASP/NTU Programme (Project No. M4082189) and the DSAIR@NTU Grant (Project No. M4082418).
\end{acknowledgments}

\bibliography{Citation}

\end{document}